%% file: Main.tex
\documentclass[sigconf]{acmart}

\AtBeginDocument{%
  }

\usepackage{booktabs}
\usepackage{svg}
\copyrightyear{2025}
\acmYear{2025}
\setcopyright{cc}
\setcctype{by}
\setcctype{by}
\acmConference[ASSETS '25]{The 27th International ACM SIGACCESS Conference
on Computers and Accessibility}{October 26--29, 2025}{Denver, CO, USA}
\acmBooktitle{The 27th International ACM SIGACCESS Conference on Computers
and Accessibility (ASSETS '25), October 26--29, 2025, Denver, CO, USA}
\acmDOI{10.1145/3663547.3746356}
\acmISBN{979-8-4007-0676-9/2025/10}
\begin{document}

\title{Understanding the Video Content Creation Journey of Creators with Sensory Impairment in Kenya}

\author{Lan Xiao}
\affiliation{%
  \institution{Global Disability Innovation Hub, UCL Interaction Centre, University College London}
  \city{London}
  \country{United Kingdom}
}
\email{l.xiao.22@ucl.ac.uk}
\orcid{0009-0008-3468-7919}

\author{Maryam Bandukda}
\affiliation{%
  \institution{Global Disability Innovation Hub, UCL Interaction Centre, University College London}
  \city{London}
  \country{United Kingdom}
}
\email{m.bandukda@ucl.ac.uk}
\orcid{0000-0002-2367-6471}

\author{Franklin Mingzhe Li}
\affiliation{%
  \institution{Human-Computer Interaction Institute, Carnegie Mellon University}
  \city{Pittsburgh}
  \country{United States}
}
\email{mingzhe2@cs.cmu.edu}
\orcid{0000-0003-4995-4545}

\author{Mark Colley}
\email{m.colley@ucl.ac.uk}
\orcid{0000-0001-5207-5029}
\affiliation{%
  \institution{UCL Interaction Centre}
  \city{London}
  \country{United Kingdom}
}

\author{Catherine Holloway}
\affiliation{%
  \institution{Global Disability Innovation Hub, UCL Interaction Centre, University College London}
  \city{London}
  \country{United Kingdom}
}
\email{c.holloway@ucl.ac.uk}
\orcid{0000-0001-7843-232X}

\renewcommand{\shortauthors}{Xiao et al.}

\begin{abstract}
Video content creation offers vital opportunities for expression and participation, yet remains largely inaccessible to creators with sensory impairments, especially in low-resource settings. We conducted interviews with 20 video creators with visual and hearing impairments in Kenya to examine their tools, challenges, and collaborative practices. Our findings show that accessibility barriers and infrastructural limitations shape video creation as a staged, collaborative process involving trusted human partners and emerging AI tools. Across workflows, creators actively negotiated agency and trust, maintaining creative control while bridging sensory gaps. We discuss the need for flexible, interdependent collaboration models, inclusive human-AI workflows, and diverse storytelling practices. This work broadens accessibility research in HCI by examining how technology and social factors intersect in low-resource contexts, suggesting ways to better support disabled creators globally.
\end{abstract}





\maketitle

\input{Chapters/1_Introduction}

\input{Chapters/2_Related_Works}

\input{Chapters/3_Method}

\input{Chapters/4_Findings}

\input{Chapters/6_Discussion}

\input{Chapters/7_Conclusion}
\begin{acks}
This project was funded with UK International Development from the UK government (Grant ID: GB-GOV-1-300815).
\end{acks}
\newpage
\bibliographystyle{ACM-Reference-Format}
\bibliography{references}

\appendix

\end{document}

%% file: Chapters/1_Introduction.tex
\section{INTRODUCTION}

Video creation has become a global medium for self-expression and communication~\cite{ha_global_2022}. However, due to its inherently multimodal nature, the video production process can pose significant barriers for creators with sensory impairments~\cite{seo_understanding_2021, cao_voices_2024}. Worldwide, an estimated 596 million individuals are blind or partially sighted (BPS)~\cite{burton_lancet_2021}, and $\approx$70 million are deaf or hard of hearing (DHH)~\cite{the_world_federation_of_the_deaf_our_nodate}. Ensuring equitable and meaningful participation for these communities remains a critical challenge. For instance, BPS creators frequently encounter difficulties with tasks such as aligning a camera or navigating interfaces primarily designed for sighted users~\cite{zhang_understanding_2023, seo_understanding_2021}. Video editing tools often lack adequate non-visual support for accessing content, evaluating quality, or independently navigating timelines~\cite{huh_avscript_2023}. Similarly, DHH creators face challenges with poor audiovisual synchronization, insufficient audio fidelity~\cite{cao_voices_2024}, and persistent shortcomings in captioning quality. This gap that has spurred grassroots advocacy, including campaigns such as \textit{\#NoMoreCraptions}~\cite{li_exploration_2022}.

While prior research has predominantly focused on making video content more accessible for audiences with sensory impairments~\cite{li_exploration_2022}, a growing body of work has shifted attention toward the experiences of disabled content creators. Recent studies have begun to unpack the practices, challenges, and needs of these creators~\cite{seo_understanding_2021, zhang_designing_2024, zhang_understanding_2023, tang_community-driven_2023, cao_voices_2024}, alongside the development of novel tools to support accessible video production~\cite{huh_avscript_2023, lee_altcanvas_2024, yoo_elmi_2025}. However, these efforts largely centre on creators in high-income contexts, overlooking the experiences of sensory-impaired creators in low- and middle-income countries (LMICs). In these settings, digital content creation is not only a form of self-expression but also a vital source of livelihood~\cite{tm_con_tmcon_2024, dzogbenuku_make_2021}, and serves as a critical avenue for advocating transparency, human rights, and social equity~\cite{simone_hinrichsen_scaling_2024, mehta_pushing_2021, li_i_2021}.

To address this gap, we examine the following research questions (RQs):

\begin{itemize} 
    \item \textbf{RQ1:} What are video content creation practices of content creators with sensory disabilities in LMICs and what challenges do they face?
    \item \textbf{RQ2:} What design considerations can inform the development of more accessible video creation tools for content creators with sensory impairments in LMICs? 
\end{itemize}
To explore the end-to-end video creation practices of disabled creators in LMIC contexts, we conducted a qualitative study with 20 video content creators in Nairobi, Kenya, 10 with visual impairments (including Blindness and Low Vision) and 10 with hearing impairments (Deaf or Hard of Hearing). Guided by the Interdependence Framework~\cite{bennett_interdependence_2018} and the Ability-Diverse Collaboration Framework~\cite{xiao_systematic_2024}, our study examined how creators navigate video production workflows, focusing on the tools and platforms they use, the genres of content they produce, and the accessibility barriers encountered across the production pipeline. Through thematic analysis of in-depth interviews, we surfaced individual strategies and collaborative practices. For example, some creators engage sighted peers to provide visual support during editing, while others work with sign language interpreters for voice-over narration.  We also investigated how participants are currently experimenting with, or envisioning, Artificial Intelligence (AI)-assisted tools such as Large Language Model (LLM), image/ video descriptions, and even video generation to support their creative processes.

Our contributions are as follows:
\begin{enumerate}
    \item We present the first in-depth qualitative study of video content creation by people with visual and hearing impairments in Kenya. By centring creators in a resource-constrained setting, we expand the geographical and sociotechnical scope of accessibility research, which has been predominantly situated in high-income regions.
    
    \item We provide empirical evidence of how creators leverage ability-diverse collaboration, reflecting the notion of "interdependence." Our study shows how creators collaborate with family, peers, and community members to overcome accessibility barriers and enhance creative control and agency.
    
    \item We identify strategies employed by creators to address systemic inaccessibility, such as role delegation, adaptive technology use, and innovative repurposing of mainstream tools. These strategies extend current understandings of ability-diverse collaboration in video content creation.
    
    \item We offer design recommendations for enhancing video editing tools and platforms, emphasizing multimodal interactions, integrated AI-driven accessibility features, and tailored remote collaboration environments. These recommendations aim to support creators with disabilities by enhancing their agency, efficiency, and expressive diversity in their creative processes.
\end{enumerate}

%% file: Chapters/2_Related_Works.tex
\section{BACKGROUND AND RELATED WORK}
Accessibility in digital content has long been a research focus, particularly concerning consumption by individuals with sensory impairments \cite{liu_what_2021, hong_video_2011, yuksel_human---loop_2020, li_recipe_2024,li_non-visual_2021,li_it_2022, li_understanding_2023}. Recent work focuses on understanding and supporting content creation by these individuals, especially video production~\cite{zhang_understanding_2023, cao_voices_2024}. This shift acknowledges the importance of inclusive participation in digital media production. In the following sections, we review studies exploring the video content creation activities of individuals with sensory impairments, the social contexts influencing these activities, and the technological interventions to facilitate accessible video content creation.

\subsection{Digital Content Creation in Low- and Middle-Income Countries}

By the end of 2021, 515 million individuals, $\approx$46\% of the population, were subscribed to mobile services in Sub-Saharan Africa~\cite{gsma_mobile_2022}, with smartphone adoption projected to reach nearly 87\% by 2030~\cite{gsma_mobile_2023}. The spread of smartphones has made video platforms important for both self-expression and earning income across the region~\cite{tm_con_tmcon_2024}. In the face of infrastructural challenges such as limited resources and funding, digital publishers in the Global South often adopt creative “platform-mixing” strategies to boost engagement and extend their reach~\cite{nielsen_born_2022}. Similar patterns have emerged in India, where content creators are leveraging these platforms not only to pursue financial stability but also to build community solidarity and catalyse social change~\cite{mehta_pushing_2021}.

Despite growing participation, content creators in LMICs continue to face significant structural barriers when compared to their counterparts in higher-income contexts. These challenges include unreliable internet connectivity, high data costs, limited access to professional editing tools, and underdeveloped monetization infrastructures~\cite{parks_internet_2020}. Navigating such constraints often requires substantial ingenuity and adaptability. For example, \citet{dzogbenuku_make_2021} describe how Ghanaian creators creatively overcome systemic obstacles to build sustainable business models. Similarly, \citet{kigundu-toure_act_2020} illustrate how socio-economic factors such as data affordability and class inequality, as well as infrastructural issues like inconsistent internet access, shape the aesthetics and workflows of South African creators, who often rely on improvised methods to maintain production and visibility. In parallel, recent reports underscore the critical need to cultivate digital skills such as creating, editing, and distributing content, as these competencies directly influence creators’ ability to participate in digital economies and enhance their income potential~\cite{simone_hinrichsen_scaling_2024, lani_developing_2021}.

Emerging research highlights the innovative use of video content creation in educational and health domains across LMICs. In particular, applying human-centred design principles, such as sustained community engagement, iterative prototyping, cultural localization, and the incorporation of relatable narratives, has shown significant promise in enhancing the effectiveness of video-based health education and improving both engagement and learning outcomes~\cite{adam_human-centered_2019}. Similarly, a recent UNICEF report on personalized learning underscores the growing importance of creator-produced educational videos that are contextually adapted to the unique pedagogical needs of learners in LMICs~\cite{unicef_trends_2022}.

Recent studies also emphasize the value of participatory and community-generated media in amplifying marginalized voices in development contexts. Saha et al. highlight the effectiveness of participatory videos created by disadvantaged women farmers in rural Bangladesh as a meaningful tool for capturing community voices, thereby facilitating direct input into international program development and enhancing community agency and empowerment~\cite{saha_community_2023}. Furthermore, research has demonstrated the potential of participatory audio methods for amplifying counter-collective narratives, providing safe spaces for marginalized communities such as women farmers in Bangladesh to authentically express and document their lived experiences and challenges~\cite{saha_hearing_2024}.

However, few studies have looked specifically at disabled creators in LMICs. Current literature predominantly examines infrastructural and economic factors broadly, neglecting specific accessibility challenges and inclusive design practices necessary for creators with disabilities. 

 \subsection{Accessibility Challenges in Video Content Creation}

Despite the revolutionary promise of digital platforms like TikTok and YouTube, many mainstream video editing tools remain predominantly designed for fully sighted and hearing users~\cite{huh_avscript_2023, cao_voices_2024}, placing creators with sensory disabilities at significant disadvantages.

For BPS creators, editing interfaces that heavily depend on visual cues, such as timelines, icons, and preview screens, pose substantial barriers, as these features are typically incompatible with assistive technologies like screen readers. \citet{zhang_understanding_2023} illustrate how such limitations impair BPS creators’ ability to fully engage in visual storytelling. These challenges are particularly pronounced in resource-constrained contexts, where BPS creators often rely on informal assistance and ad hoc solutions to navigate inaccessible editing tools~\cite{vashistha_educational_2014}. Additionally, mainstream software frequently lacks interfaces supporting non-visual interaction, significantly marginalizing BPS creators who seek to produce high-quality content~\cite{radanliev_accessibility_2024}.

Beyond technical constraints, BPS creators face significant barriers to social engagement on video-sharing platforms. \citet{seo_understanding_2021} identify that while blind vloggers actively advocate for disability awareness through platforms such as YouTube, they continually struggle with visual interactions and meaningful participation in digital dialogues. To address these limitations, novel assistive tools like \textit{AVscript} offer a text-based editing environment optimized for screen reader navigation to allow blind creators to embed visual descriptions and correct common video-production errors~\cite{huh_avscript_2023}. Similarly, \textit{AltCanvas} employs generative AI to provide a dynamic tile-based interface for composing complex visual narratives, reducing reliance on traditional visual affordances~\cite{lee_altcanvas_2024}.

Creators who are Deaf and DHH encounter distinct challenges due to the inherently audio-centric nature of video production workflows. Current editing environments typically lack sufficient support for caption synchronization and sign language integration, significantly limiting expressive capabilities~\cite{kuksenok_accessible_2013}. Furthermore, automated captioning tools often generate inaccurate or contextually inappropriate outputs, requiring extensive manual corrections and thus compounding labour for DHH creators \cite{li_exploration_2022, alonzo_beyond_2022}. Recent analyses by Cao et al. of Deaf-authored TikTok content highlight the difficulty of aligning visual storytelling effectively with sign language, captions, and audio elements, underscoring the need for improved translation and editing tools \cite{cao_voices_2024, tang_community-driven_2023}.

In addition to infrastructural and economic constraints, social and algorithmic barriers further limit the visibility and impact of content produced by disabled creators. Research points to persistent issues of stigma, systemic neglect, and a lack of institutional support~\cite{niu_please_2024}. In response, creators such as blind TikTokers have adopted strategic practices to maintain content quality and visibility, despite platform limitations~\cite{lyu_because_2024}. However, recommendation algorithms often exacerbate exclusion by failing to account for the unique needs and content patterns of disabled creators, leading to experiences of disappointment and marginalization~\cite{choi_its_2022}. Additionally, automated content moderation systems may misclassify responses to harassment, inadvertently penalizing disabled users and reinforcing inequities in platform governance~\cite{lyu_i_2024}.

\subsection{Ability-Diverse Collaboration and Interdependence Among Creators with Sensory Impairments}

Emerging research in assistive technology design emphasizes two complementary frameworks that challenge the traditional emphasis on individual independence. First, interdependence models highlight the benefits of collaboration and shared contributions among creators with sensory impairments. \citet{bennett_interdependence_2018} introduce interdependence as a model in which assistive technologies develop from mutual support among users, while \citet{bostad_freedom_2016} argue for embracing interconnected dimensions of independence, dependence, and interdependence to achieve greater freedom for disabled individuals. Second, \citet{xiao_systematic_2024} present a systematic review of ability-diverse collaboration in Human-Computer Interaction (HCI), introducing the \textit{Ability-Diverse Collaboration Framework}. This framework articulates two modes of interdependent interaction: ability sharing and ability combining. It highlights how structured, role-sensitive collaboration between individuals with differing abilities can reduce asymmetries in information access, foster equitable participation, and support more inclusive technology design.

These frameworks work in tandem to provide an approach to inclusive technology design. Ability-diverse collaboration serves as the primary analytical lens for understanding how different capabilities can be strategically combined and shared in collaborative settings. Meanwhile, interdependence functions as a critical tool for surfacing and addressing the relational and structural inequalities that often emerge in collaborative environments. Where ability-diverse collaboration provides the operational framework for organizing collaborative interactions, interdependence offers the theoretical foundation for recognizing how power dynamics, resource access, and social structures influence collaborative outcomes. Together, they enable researchers and designers to both facilitate meaningful collaboration and critically examine the systemic barriers that may limit equitable participation.

Empirical research increasingly supports the value of these interdependent frameworks, demonstrating how collaborative approaches can drive innovation and inclusion in technology and art. Hou's work on co-creating multisensory art with blind and visually impaired participants highlights how tactile, auditory, and olfactory modalities, combined with personalized colour-encoding systems, facilitate shared aesthetic experiences between blind and sighted audiences\cite{hou_making_2023}. \citet{hendriks_codesign_2015} argue for the use of "method stories" as a reflective co-design practice, emphasizing that standardized approaches often fail to accommodate the diverse sensory and communicative preferences of participants with cognitive or sensory impairments. Similarly, Brulé et al.'s \textit{MapSense} system exemplifies interdependent design through the integration of tactile, auditory, and haptic feedback to support spatial learning for visually impaired children\cite{brule_mapsense_2016}. \citet{barbareschi_speech_2024} document how nuanced micro-communications, such as vocal cues and bodily gestures, foster mutual coordination and trust between blind runners and their sighted guides. Cooperative task engagement has also been shown to enhance both performance and satisfaction among individuals with multiple disabilities~\cite{lancioni_engagement_2002}.

On a technological level, \citet{teng_help_2024} demonstrate how assistive systems can facilitate face-to-face collaboration between blind and sighted individuals in digital media production, expanding practical interdependence. \citet{das_simphony_2023} introduced \textit{Simphony}, an audio-tactile system that supports collaborative pattern creation between blind weavers and sighted instructors, and developed \textit{Co11ab}, a Google Docs extension that integrates speech and non-speech auditory cues to support real-time co-writing by screen reader users and sighted collaborators~\cite{das_co11ab_2022}. Collectively, these studies illustrate that adopting interdependent practices not only improves accessibility and creative output but also meaningfully empowers creators with sensory disabilities.

%% file: Chapters/3_Method.tex
\section{METHOD}
To investigate the video creation experiences of sensory-impaired creators living in LMICs, we conducted a qualitative study involving semi-structured interviews with video content creators in Kenya. The study received approval from the university’s ethics committee. Semi-structured interviews were selected to enable in-depth exploration of participants' practices, challenges, and adaptive strategies in video production. All interviews were conducted in accordance with established ethical guidelines, with participants providing informed consent and their confidentiality fully protected.

\subsection{Participants}
We recruited 20 video content creators residing in Kenya, comprising 10 individuals who are BPS and 10 who are DHH. To be eligible, participants were required to self-identify as having a visual or hearing impairment and to have at least six months of experience in video content creation. Participants ranged in age from 21 to 42 years (M = 25.75, SD = 5.05), with a balanced gender distribution across groups. Recruitment was conducted in collaboration with a local organization that supports individuals with sensory impairments.

The BPS group included six participants who are completely blind and four with low vision, consisting of six males and four females. Their professional roles demonstrated strong involvement in creative and educational sectors: three were professional musicians (P2, P8, P9), two were educators (P3, P10), and five were students (P10, P12, P13, P17, P19). In terms of content creation experience, four participants reported more than three years (P2, P8, P9), while six had between one and three years of experience (P3, P10, P12, P13, P17, P19), while none reported less than one year of experience. 

The DHH group comprised six individuals who are deaf and four who are hard of hearing, with a gender distribution of four males and six females. Their occupational backgrounds were equally varied, spanning roles such as filmmakers (P1, P4), dancer (P5), marketer (P6), comedian (P7), and sign language interpreter (P15). Additionally, several participants were students (P11, P14, P18), reflecting a younger demographic engaged in content creation alongside their education. Four participants had more than three years of experience (P6, P7, P16), three had between one and three years (P11, P14, P15), and three had less than one year (P1, P4, P5).  \autoref{tab:demo} provides detailed demographic information.
\begin{table*}[ht]
\begin{tabular}{@{}llllll@{}}
\toprule
\textbf{ID} & \textbf{Age} & \textbf{Gender} & \textbf{Disability} & \textbf{Occupation}        & \textbf{Content Creation Experience} \\ \midrule
P1          & 23           & Male            & Deaf                & Filmmaker                  & Less than one year                   \\
P2          & 29           & Male            & Blind               & Musician                   & More than 3 years                    \\
P3          & 26           & Female          & Blind               & Teacher                    & 1-3 years                            \\
P4& 24           & Male            & Hard of Hearing     & Filmmaker                  & Less than one year                   \\
P5          & 24           & Female          & Deaf                & Dancer                     & Less than one year                   \\
P6          & 24           & Female          & Deaf                & Marketing                  & More than 3 years                    \\
P7          & 42           & Male            & Deaf                & Comedian                   & More than 3 years                    \\
P8          & 25           & Male            & Blind               & Musician                   & More than 3 years                    \\
P9          & 30           & Male            & Low Vision          & Musician                   & 1-3 years                            \\
P10         & 26           & Female          & Blind               & Student                    & 1-3 years                            \\
P11         & 22           & Female          & Hard of Hearing     & Student                    & 1-3 years                            \\
P12         & 22           & Male            & Blind               & Student                    & 1-3 years                            \\
P13         & 24           & Male            & Blind               & Student                    & 1-3 years                            \\
P14         & 24           & Female          & Deaf                & Student                    & 1-3 years                            \\
P15         & 24           & Female          & Deaf                & TV Sign Language Interpter & 1-3 years                            \\
P16         & 35           & Male            & Deaf                & Musician                   & More than 3 years                    \\
P17         & 22           & Female          & Low Vision          & Social Worker              & 1-3 years                            \\
P18         & 22           & Female          & Deaf                & Student                    & 1-3 years                            \\
P19         & 25           & Female          & Low Vision          & Student                    & 1-3 years                            \\
P20         & 21           & Male            & Low Vision          & Student                    & 1-3 years                            \\ \bottomrule
\end{tabular}
\caption{Participants’ demographic information and content creation experience}
\label{tab:demo}
\end{table*}
\subsection{Procedure}

The semi-structured interviews lasted 45 to 60 minutes and were carried out online using Zoom. For participants with hearing impairments, Sign language interpretation was provided as needed. The interview protocol encompassed four main sections:

\begin{itemize}
    \item Demographic Information: Participants were asked about their age, gender, occupation, type and degree of sensory impairment, and the assistive technologies they use.
    \item Video Content Creation Practices and Challenges: We explored participants' typical workflows, including the devices and software they use. We also delved into their challenges during the content creation process, such as technical difficulties, accessibility barriers, and resource limitations.
    \item Collaborative Strategies and Social Interactions: Participants were asked about their experiences collaborating with others during content creation, including the roles of collaborators, the nature of interactions, and any concerns related to collaboration. We also inquired about their participation in content-creator communities and support networks.
    \item Use of AI Tools: We investigated participants' awareness and use of AI-based tools in their video production processes, discussing perceived benefits, limitations, and areas for improvement.
\end{itemize}

Interviews were audio-recorded with participants' consent and transcribed verbatim for analysis.

\subsection{Data Analysis}
The interview data were analysed using a hybrid thematic analysis approach~\cite{braun_reflecting_2019}, combining inductive and deductive strategies informed by the research questions. Initial coding was conducted by the first author, who read the interview transcripts line by line and employed open coding techniques at the sentence and paragraph levels to identify meaningful insights. This process was iterative, involving multiple rounds of refinement and consolidation to ensure analytic depth. In the early stages, coding focused on identifying tools, practices, and challenges encountered by sensory-impaired creators during video production. Building on these initial codes, we then examined how participants engaged in collaborative practices and social interactions throughout the creative process. Related codes were grouped to develop broader conceptual categories and candidate themes. Following the development of initial themes, we revisited the transcripts to assess their alignment with the data, examining instances that both supported and challenged each theme. Throughout the analysis, five co-authors with expertise in qualitative research participated in reviewing and refining the subtheme categorizations during team meetings. Through this collaborative process, we arrived at three final overarching themes: (1) Tools for Video Content Creation,  (2) Challenges as Content Creator with Sensory Impairment, and (3) Workflow and Collaborative Strategies in Video Content Creation.

%% file: Chapters/4_Findings.tex
\section{FINDINGS}

In this section, we present the interview findings that highlight the participants' content creation practices, the challenges they faced, and the strategies they used to create their optimum video content creation journey. Our analysis revealed key insights related to the specific tools and techniques used by creators, accessibility barriers encountered, collaboration practices, and the role of AI in facilitating content creation. We organize our findings into 3 thematic subsections: Tools for Video Content Creation, Challenges as Creators with Sensory Impairment, and Workflow and Collaborative Strategies in Video Content Creation to illustrate the complex interactions between creators' impairments, available resources, and their social contexts.

\subsection{Tools for Video Content Creation}
Creators with sensory impairments employ various hardware and software tools to create video content. Most of the participants (15 out of 20) mainly used their smartphones to complete the entire production workflow - filming, editing, and sharing. A smaller subset supplemented their setup with additional devices: three participants used stand-alone cameras for filming (P6(DHH), P9(BPS), and P16(DHH)), and another three used laptops specifically for editing (P1(DHH), P6(DHH), and P8(BPS)).

All participants used mobile phones for video creation. To better understand their production capabilities, we enquired about the specific phone models in use. Most reported using entry-level to mid-range smartphones (e.g., Samsung Galaxy A14, HMD M-KOPA X20), typically priced between \$100 and \$200. Only three participants used newer mid-to-upper-range devices (e.g., iPhone 11, Oppo Reno 13), with prices ranging from \$300 to \$500.

Regarding software, most participants used the default camera application on their phones for video recording. Only two (P1(BPS) and P8(BPS)) reported using alternative apps, such as \textit{Blackmagic Camera} and \textit{Open Camera}. For editing, 12 participants used \textit{TikTok}’s built-in editing features, while nine used the companion app \textit{CapCut} in conjunction with \textit{TikTok}. Other editing tools mentioned included \textit{InShot}, \textit{BPSDs}, and more advanced software such as \textit{Adobe Premiere}, \textit{DaVinci Resolve}, and \textit{Lightroom}. Since 4 participants were music professionals, audio editing tools such as \textit{Logic Pro}, \textit{Cubase}, and \textit{Virtual DJ} were also reported.

\begin{figure*}
    \centering
    \includegraphics[width=0.8\linewidth]{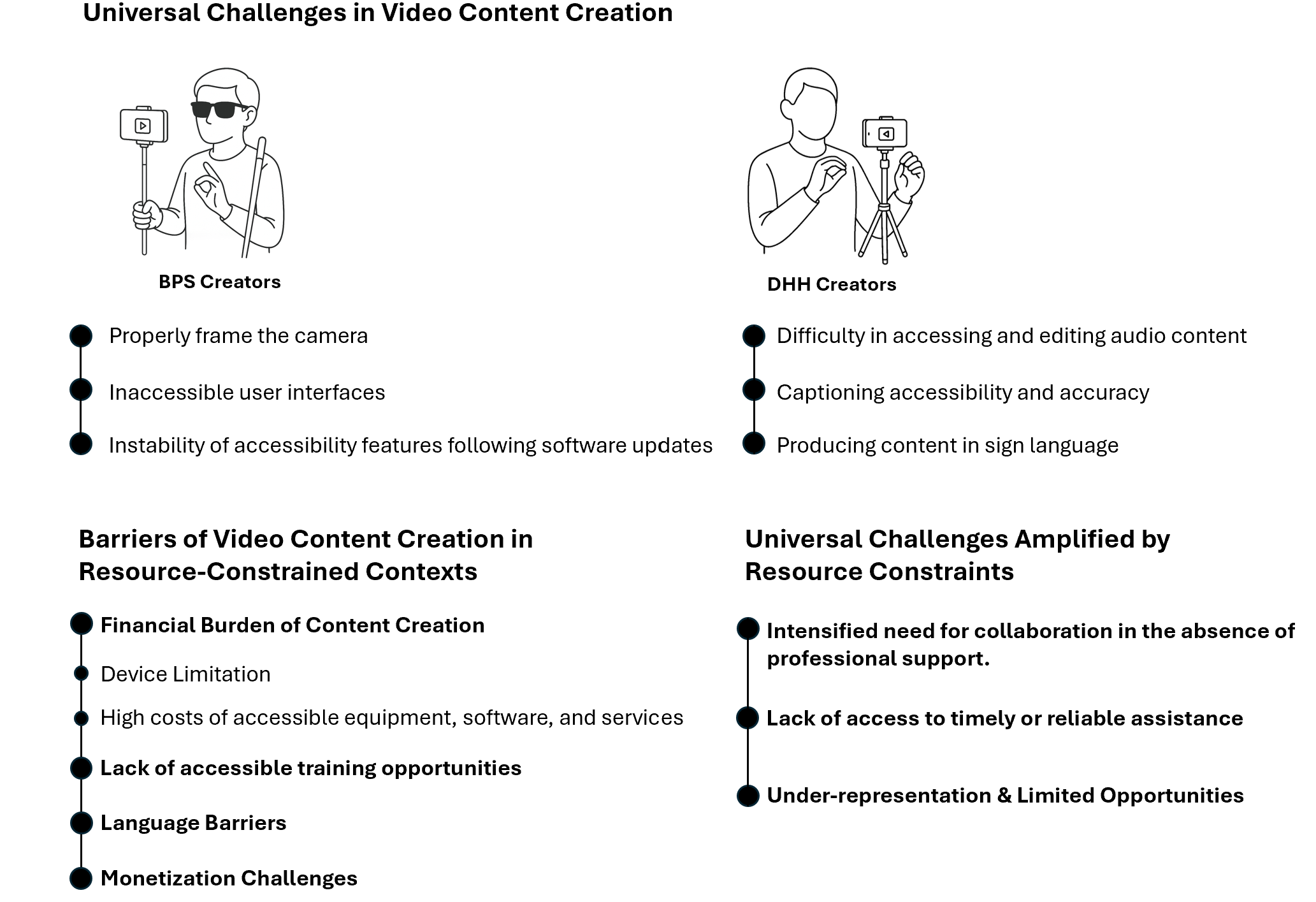}
    \caption{Challenges as Creators with Sensory Impairments}
    \Description{This infographic illustrates the barriers faced by content creators with sensory impairments. For BPS (Blind and Partially Sighted) creators, challenges include difficulties in properly framing the camera, navigating inaccessible user interfaces, and dealing with unstable accessibility features following software updates. DHH (Deaf and Hard of Hearing) creators face issues such as difficulty accessing and editing audio content, captioning accessibility and accuracy problems, and producing content in sign language. Beyond these creator-specific challenges, the infographic identifies barriers in resource-constrained contexts including the financial burden of content creation, device limitations, high costs of accessible equipment and software, lack of accessible training opportunities, language barriers, and monetization challenges. The diagram further shows how resource constraints amplify universal challenges, leading to an intensified need for collaboration without professional support, lack of access to timely or reliable assistance, and under-representation with limited opportunities for creators with disabilities.}
    \label{fig:challenges}
\end{figure*}
\subsection{Challenges as Creators with Sensory Impairment}

Participants reported a range of challenges throughout their video content creation journey. We present these findings in two main sections that highlight the unique context of content creation in resource-constrained environments: first, we examine barriers specific to LMIC contexts, then explore how universal accessibility challenges are amplified by resource constraints.

\subsubsection{Barriers of Video Content Creation in Resource-Constrained Contexts}

\textbf{Financial barriers to accessing necessary support services and tools} posed substantial challenges unique to resource-constrained environments. P15 (DHH) explained how essential services became financially prohibitive: "\textit{Communication is a challenge...also asking for people to come and help is very hard. For example, interpreters, you must pay them.}" This financial burden extended to software and specialized tools. P6 (DHH) described the limitations of using free versions of editing apps: "\textit{You need to pay for some aspects... I think it's easier when I use the free one on my laptop.}" P14 (DHH) emphasized the impact on content quality: "\textit{There's some technology or software, especially for lighting and sound, that requires payment. So I think when I have them, I can improve the quality of my videos.}" Even basic accessibility features often required paid subscriptions, as P16 (DHH) noted: "\textit{Now with the hearing accessibility features...that requires payment. That is where I have a problem.}"

Participants across both groups faced significant challenges related to \textbf{device limitations} that directly impacted their content quality and accessibility. These limitations were particularly acute given the financial constraints of acquiring newer technology. P7 (DHH) shared: "\textit{My phone is poor. I cannot include both signing and captioning… the one I'm using has no captioning feature.}" 

The \textbf{lack of accessible training opportunities} emerged as a critical barrier, with only 4 out of 20 participants having received any formal training in content creation. This gap meant that most creators were self-taught, learning through trial and error without the benefit of structured guidance or accessibility-specific instruction. The absence of formal training programs tailored to creators with sensory impairments in local contexts left many struggling to develop professional-level skills independently.

\textbf{Language barriers presented unique challenges in the African context}, where creators needed to serve multilingual audiences but lacked adequate support. P15 (DHH) explained: "\textit{Even in wording...using the right language or the right sentences...it's a challenge because I don't understand English perfectly. Most of my viewers prefer Swahili or other languages. Even when captions are available in English, my viewers struggle to understand them. So it becomes a challenge for them as well.}" This gap in local language support meant that even when accessibility features like auto-captioning were available, they often failed to serve the actual linguistic needs of both creators and their audiences in African contexts.

The \textbf{structural disparities in platform ecosystems particularly disadvantaged African creators} with sensory impairments. P6 (DHH) highlighted how disability intersected with platform algorithms to limit earning potential: "\textit{For you to monetize your content, if you don't have enough followers, or if you don't have voicing to attract attention, it's difficult to earn from it.}" P18 (DHH) explained the audience mismatch: "\textit{Most of the viewers are hearing people. They will not be able to get my point… Not so many deaf people are able to get my content.}" These monetization challenges were compounded by limited platform support for African creators and fewer opportunities for brand partnerships compared to creators in higher-resource contexts.

\subsubsection{Universal Challenges Amplified by Resource Constraints}

Participants encountered several universal accessibility challenges that align with existing literature on creators with sensory impairments \cite{zhang_understanding_2023, seo_understanding_2021}, including \textbf{inaccessible user interfaces}, \textbf{absence of accurate captioning tools}, and \textbf{discrimination on platforms} (see \autoref{fig:challenges} for comprehensive details).  P8 (BPS) noted unlabelled buttons requiring sighted assistance, while P20 (BPS) described apps becoming unusable in editing features. Similarly, P18 (DHH) highlighted missing captions as barriers to understanding content, and auto-captioning inaccuracy was reported by multiple participants. Discrimination manifested through exclusion and bullying, as P17 (BPS) shared: "\textit{They used to say… you are not fit for this account.}" 

While these challenges exist globally, the resource-constrained context particularly amplified other barriers. The \textbf{need for assistance in content creation became more critical} in the absence of professional support services. P13 (BPS) captured this dependency: "\textit{The biggest challenge is that you cannot do it by yourself... You require some other person to be there... which is not often the case.}" A similar need emerged in creative contexts for DHH creators, where the lack of accessible audio feedback made independent editing difficult. As P14 (DHH) explained: "\textit{When I want to copy a dancing video from someone else, I can't hear the sound, not even the vibration. So, I'm only able to replicate the dance moves, but I can't listen to the song. I always need someone to help me match the song and the dance together.}"

This dependence on assistance was further intensified by device limitations. P10 (BPS) noted their reliance on others due to hardware constraints: "\textit{Most of the time I rely on someone to help me out… since my phone has not yet upgraded.}" Communication barriers were amplified when working across ability differences without access to interpretation services. P18 (DHH) elaborated: "\textit{Communication becomes hard, especially when collaborators are hearing and they don't understand the language.}" P7 (DHH) added: "\textit{Getting somebody to interpret exactly what I want, most of the time I get stuck... so I decided to focus more with deaf people.}"

The reliability of informal support networks also posed challenges. P15 (DHH) shared: "\textit{I have one customer who helped me for a year, but she travelled. She's no longer here... I haven't gotten anybody this year.}" This dependence on informal, unpaid assistance meant that production schedules were often at the mercy of others' availability. P10 (BPS) recounted how strangers were skeptical of her due to her appearance and assistive devices: "\textit{They think I am lying... convincing someone at some point becomes difficult, especially someone who is a stranger.}"

The \textbf{structural disparities between creators with and without disabilities were particularly pronounced} in resource-constrained contexts. P16 (DHH) explained: "\textit{Deaf people have a lot of challenges in content creation compared to hearing counterparts… They have a lot of linkages and opportunities, but they don't want to link with people who are deaf.}" Without institutional support or advocacy organizations to combat discrimination, individual creators bore the full burden of navigating hostile platform environments. This was compounded by limited access to safe filming locations and equipment, as P1 (DHH) noted barriers related to accessing secure spaces, while P6 (DHH) shared: "\textit{I can't get help from people who are not concerned about me or about any other deaf creators.}"

\subsection{Workflow and Collaborative Strategies in Video Content Creation }

Our interviews reveal that for creators with sensory impairments, video production is not an isolated or purely individual endeavour, but rather a fundamentally collaborative and staged process. Therefore, it was not possible in the analysis to separate out the collaboration from the processes. Participants consistently described their creative workflows as shaped by the interplay between sensory access constraints, available assistive technologies, and the roles of trusted collaborators. Across both groups, DHH and BPS creators, we identified five shared stages: \textbf{pre-production}, \textbf{filming}, \textbf{visual editing}, \textbf{sound editing}, and \textbf{post-production}. While collaboration was a common thread across all stages, the nature of collaborators’ involvement, the forms of dependency, and the modalities of engagement varied significantly between groups.

\begin{figure*}[t!]
\centering
\includegraphics[width=1\linewidth]{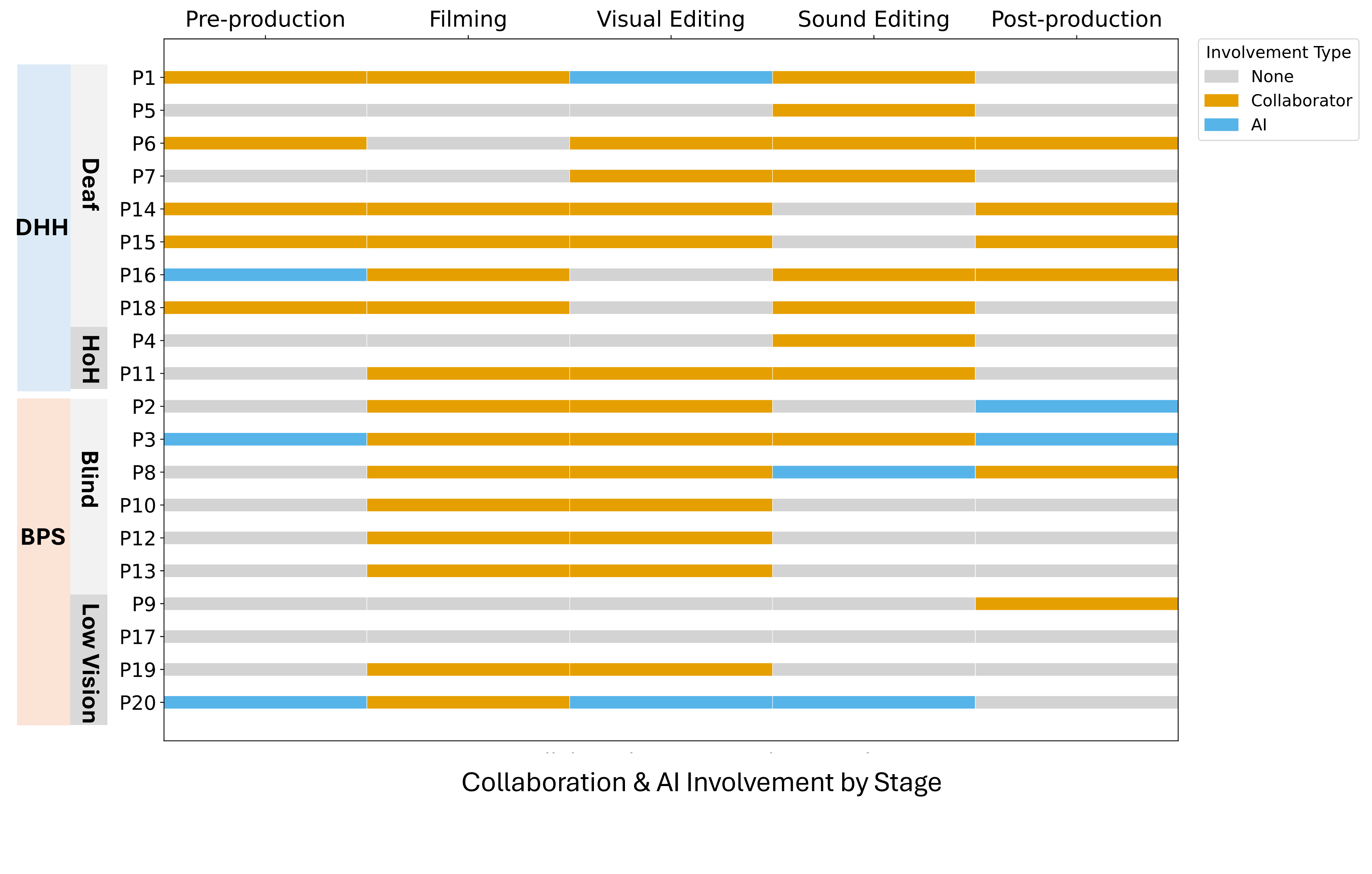}
\caption{Collaboration and AI Involvement Across Stages of Video Production for Content Creators with Sensory Impairment. This matrix visualizes the involvement of collaborators and AI tools across five stages: Pre-production, Filming, Visual Editing, Sound Editing, and Post-production, for 20 content creators with various disabilities. Each row represents a participant, and each coloured cell indicates if a collaborator (orange) or AI tool (blue) was involved at that stage. Gray indicates no external support or creators chose to skip that stage for some reason.}
\label{fig: collaboration}
\Description{A matrix chart showing the involvement of collaborators and AI tools in five stages of video production: Pre-production, Filming, Visual Editing, Sound Editing, and Post-production, for 20 content creators with sensory impairments. The chart is organized by disability groups: Deaf, Hard of Hearing (HoH), Blind, and Low Vision. Each row represents a participant (P1 to P20). Each cell is colour-coded: orange for collaborator involvement, blue for AI use, and gray for no involvement or stage skipped. Most DDH participants (P1–P7, P14–P16, P18) involve collaborators in multiple stages, especially during Filming and Editing. AI involvement is more common among BPS participants (e.g., P2, P3, P8, P20), particularly in Visual and Sound Editing. Some participants (e.g., P4, P10, P13) have mostly gray cells, indicating limited or no external support. A legend in the top right clarifies the colour coding. }
\end{figure*}

Collaboration was nearly ubiquitous: 19 of the 20 participants, excluding P17 (BPS), who identified as having low vision, reported engaging collaborators during one or more stages of the process. These collaborators included peers, friends, interpreters, sighted assistants, and increasingly, AI-driven tools. Notably, six participants integrated generative or assistive AI (e.g., for scripting, sound mixing, or visual description), highlighting an emerging convergence of human and algorithmic collaboration in accessible content creation (refer to \autoref{fig: collaboration}).

To unpack the complexity of these collaborative workflows, we first present a comparative overview of the five key stages of video production, highlighting points of convergence and divergence in how collaboration and AI usage unfolds across DHH and BPS creators. We then examine how agency, trust, and concerns are negotiated in content creation, focusing on creators' interactions with both human collaborators and AI technologies.

\subsubsection{Pre-production: Planning, Translation, and Visualization }
In the pre-production stage, creators from both groups initiated their creative vision independently but relied on distinct forms of collaborative support to navigate sensory access constraints.

DHH creators often co-developed scripts or conceptualized video ideas in dialogue with other DHH peers. When content involved music or spoken dialogue, hearing collaborators and sign language interpreters played critical roles in translating lyrics or audio elements into sign language and supporting rhythmic interpretation. For example, P5 (DHH) shared: “\textit{When I want to create content, I invite a friend. I might share the song I plan to dance to... My friend then helps by translating the lyrics into sign language for me.}” Similarly, P16 explained: “\textit{The lyrics are my idea, but when it comes to the rhythm, the beat, I need someone to help me.}”

In contrast, BPS creators frequently developed narrative arcs and scripts independently, drawing on assistive technologies such as screen readers or generative AI (e.g., large language models) to support ideation. As P3 (BPS) noted: “\textit{When I write a short introduction of my script, then I command [Gemini] to help me come up with beautiful titles.}” Notably, some DHH creators also adopted AI tools for co-authoring content; P16, for instance, reported using ChatGPT to collaboratively generate lyrics and narrative structures.

\subsubsection{Filming: Spatial Orientation and Rhythm Synchronization}
During filming, both groups engaged collaborators to address different sensory access needs, with support focused on either spatial awareness or temporal coordination.

DHH creators were generally able to operate cameras independently but sought assistance when appearing on screen, such as for travel vlogs or dialogue-based content, or when synchronizing signed performance with audio tracks. Visual cues provided by collaborators enabled more accurate alignment with musical rhythm. As P11 (DHH) described: “\textit{If it’s slow, I sign slowly. If it’s fast, I adjust my signing so that it links with the video.}” In interview settings, interpreters often facilitated real-time translation between spoken and signed language.

For BPS creators, filming centred on spatial orientation, alignment, and composition. While some leveraged residual vision or tripod setups, many incorporated assistive AI tools like Seeing AI or Google Lookout. Others relied on verbal guidance from collaborators to ensure correct positioning. As P10 (BPS) recounted: “\textit{I tell them to speak so that I can face the right side... and again I tell them to help me know if at all I have positioned myself well.}” Collaborators also supported aesthetic framing. P2 (BPS), for instance, explained: “\textit{I took videos in a place where I know I need to change the background. I normally tell them: make this background maybe... put a copy or a grass carpet.}”

\subsubsection{Visual Editing: Captioning, Composition, and Descriptive Review}
In the visual editing stage, creators with sensory impairments engaged in collaborative workflows that foregrounded distributed labour, communicative negotiation, and creative direction. While the division of roles varied between groups, most participants described visual editing as a phase that required sustained support from sighted or hearing collaborators.

DHH creators commonly managed core editing tasks such as trimming, arranging clips, and inserting transitions. However, they faced persistent barriers when working with audio elements or environmental cues that could not be perceived directly. In particular, captioning posed significant challenges, not only in transcribing spoken language but also in describing non-verbal sounds. As P6 (DHH) explained, “\textit{When I’m talking about the sound of a dog barking, how do you explain that? How do you spell that sound? For a deaf person, it’s not easy… But when you [hearing person] explains it, it becomes easier.}” These collaborations often involved interpretive labour from hearing friends or peers, who helped translate auditory scenes into meaningful visual or textual representations.

BPS creators, by contrast, typically delegated the entire visual editing process to sighted collaborators. However, this delegation did not entail passivity. Many creators provided highly specific, directive input regarding visual aesthetics, on-screen appearance, and scene composition. These instructions reflected both strategic self-presentation and a desire for visual quality aligned with their creative intent. As P2 (BPS) shared, “\textit{I tell them to remove the background and put this and this… like, for example, I want to look like I was in a studio.}” P3 (BPS) similarly expressed, “\textit{Make my face brighter, you know… or make me look younger. Or I want to be white, make me white.}”

\subsubsection{Sound Editing: Control and Exclusion }
Sound editing revealed the most pronounced divergence between groups, with DHH creators facing significant barriers and BPS creators demonstrating high levels of autonomy.

For many DHH creators, this stage was perceived as the least accessible. Tasks such as voice-over production and audio mixing were frequently outsourced or omitted altogether (e.g., by P14 (DHH) and P15 (DHH)). While some engaged with music through tactile feedback (e.g., vibrations), decisions around emotional tone or timing were often validated by hearing collaborators.

By contrast, BPS creators, especially those with musical backgrounds, exhibited strong control over sound editing. As P8 (BPS) explained: “\textit{I do all the sound by myself. I compose, I sing, I even edit the track before putting it in the video.}” AI tools such as DJ.Studio were also employed for mixing and mastering. Still, screen reader limitations occasionally necessitated support, particularly when adjusting audio settings. P10 (BPS) noted: “\textit{I usually ask a friend to check if the music starts at the right point.}” P8 (BPS) added: “\textit{To adjust the volume, like the knobs or tabs... I need a sighted person to help. Many times the screen reader doesn’t speak everything out.}”

\subsubsection{Post-Production: Multimodal Quality Checks and Platform Management}
The final stage involved multimodal review, export, and online distribution, often blending self-review, collaborator support, and AI-based quality control.

DHH creators enlisted collaborators to audit caption timing and assess overall audio quality. While visual editing and content structuring could be completed independently, verifying elements such as background noise or voice clarity often required input from hearing individuals to meet audience expectations.

BPS creators balanced self-review with trusted feedback loops, taking an active role in ensuring that final outputs met both personal and platform-specific standards. P2 (BPS) shared: “\textit{If it’s not what I wanted, I tell them to redo it... maybe they made it one minute and 50 seconds, so it means some parts were left out.}” P9 (BPS) emphasized the importance of platform literacy: “\textit{TikTok and YouTube accounts have restrictions and rules... I have to check on all that, or have someone help me check before I upload.}”

Support extended beyond content production to audience interaction. Due to screen reader limitations, BPS creators frequently involved collaborators in managing engagement. P13 (BPS) described: “\textit{They help in shooting me, training me, reading for me the comments… I involve people who can see to respond to the comments and keep the channel growing.}”

AI tools also supported quality assessment. P2 (BPS) used Seeing AI to review clothing and background elements for visual consistency, while P3 (BPS) relied on the app's description features to evaluate visual edits: “\textit{Seeing AI helps you identify your video… even the basic version helps you recognize the areas that have been edited.}”

\subsubsection{Agency, Trust, and Concern in Collaboration}
For creators with sensory impairments, video production extends beyond a series of technical tasks; It becomes an ongoing negotiation of agency, dependence, and authorship. Rather than aspiring toward full autonomy or defaulting to external assistance, participants described a form of \textit{negotiated agency}, fluidly shifting between the roles of director, collaborator, and editor in response to the task at hand, their personal preferences for privacy and control, and the trustworthiness of available support systems. This dynamic was shaped not only by the division of labour but also by the social and emotional stakes of collaboration, whether with human partners or AI technologies. 

\textbf{Collaboration as Relational Infrastructure} 
For many creators, collaboration was deeply embedded in relational contexts. Human support was not only functional but affective, often facilitated through close-knit, informal networks. Friends were the most frequently mentioned collaborators (16 participants), followed by interpreters, family members, and romantic partners. Among DHH creators, sign language interpreters played a key role in mediating cross-modal communication, while Deaf peers were preferred collaborators for co-creation due to shared lived experiences. As P6 (DHH) shared, \textit{“We’ve experienced the same challenges, so it’s easy because of the interaction with them.”}

BPS creators described similarly diverse ecosystems of support. Some worked closely with family members or classmates, while others described ad hoc assistance from strangers or fans during public filming or livestreams. As P10 (BPS) noted, \textit{“If I’m on the road, I might just bump into someone and ask them to help me take a video, it’s just once in a while.”} These interactions were often informal, built on interpersonal trust rather than formal agreements.

However, collaboration also introduced vulnerability. Several participants expressed discomfort with repeated reliance on others, especially when it risked overburdening friends or compromising creative intent. As P19 (BPS) reflected, \textit{“This is my content… and it is me who knows what I want. So calling my friend… sometimes them feeling disturbed or just getting tired, I feel bothered.”} Others described concerns around misrepresentation in tasks they could not personally verify. P12 (BPS) noted: \textit{“If I want someone to record me on a white background… and they bring me to a yellow one. I may not know the difference. I could easily be deceived.”}

Participants also raised privacy and data security concerns, particularly when relying on remote support services. For instance, P13 (BPS) questioned the supposed autonomy offered by visual support apps: \textit{“It’s really not different from calling someone… it doesn't give you independence… you still rely on someone else.”} She continued, \textit{“You have to expose your whole house to this person. You don’t know them… and that is very, very crucial when it comes to me.”} Her concerns extended to potential data breaches: \textit{“This person can decide to either screenshot the screen or do some other thing… you don’t know, people out there are not trustworthy.”}

\textbf{AI as a Tool for Control and Efficiency} 
In contrast to the social weight of human collaboration, AI tools were described as offering increased privacy, efficiency, and control. Twelve participants described using AI in daily life, and six (four BPS and two DHH) incorporated AI directly into video production. Use cases included large language models (e.g., ChatGPT, Gemini) for scripting and captioning, vision-based apps (e.g., Seeing AI, Google Lookout) for scene description, and editing platforms (e.g., DJ.Studio, Filmora, InVideo AI) for automating audio and video workflows. 

One illustrative case was P20, a BPS creator who employed multiple AI tools for content marketing. \textit{“I just write my prompt to ChatGPT: I am looking to market an electric kettle… I mention the brand, the capacity, the material… then I go to InVideo AI, paste it there, and tell it to generate the video.”} Afterward, he customized the visuals: \textit{“I always get rid of the video part and put what I have already recorded… but that’s where the challenge comes in, those videos always have watermarks.”} Since high-quality photography remained a challenge, he relied on a sighted assistant to take product photos and then used Background Eraser to isolate the object: \textit{“I just get rid of the background to remain with a white background and the product alone.”} For captions, he used generative prompts: \textit{“I just write the prompt and ask ChatGPT to generate captions for an image, then I attach that to the photos I’ve taken.”}

While AI offered a sense of control and creative efficiency, participants remained critical of its limitations. Some emphasized the need for human oversight when nuance, visual judgment, or emotional tone were involved. As P13 (BPS) stated, \textit{“I would rather ask someone I know to help me, or have some kind of AI that I can manipulate on my own.”} Even confident users like P2 (BPS) acknowledged constraints: \textit{“I still need someone to check the visual part, I can’t do everything myself.”}

Nevertheless, AI was often preferred when speed, privacy, or scheduling constraints took precedence. As P2 (BPS) explained, \textit{“When I send my videographer to edit it for me, he might take a whole day… I’ll lose the audience I’m used to posting for at 4 p.m.”} Here, AI allowed creators to better manage their schedules and meet personal deadlines without relying on others or compromising privacy.

%% file: Chapters/6_Discussion.tex
\section{DISCUSSION}
Our findings highlight the experiences and challenges faced by video content creators with sensory impairments in Kenya. Extending prior work on accessible content creation \cite{seo_understanding_2021, zhang_understanding_2023, cao_voices_2024}, this study shifts the focus beyond individual tools to the broader, often collaborative, practices that underpin video content generation. We highlight how creators’ workflows are embedded in interdependent arrangements and call attention to the infrastructural and social conditions that shape these dynamics. Below, we contextualise these findings within existing literature and present implications for designing more inclusive video production tools.

\subsection{Supporting Interdependent Creation}
HCI research on accessibility has traditionally focused on optimizing tools for individual users \cite{bennett_interdependence_2018}. Yet, our findings reveal that for video creators with hearing or visual impairments, accessibility is not solely achieved through personal adaptations. Instead, it often emerges through collaborative, interdependent practices. Family members, peers, and collaborators routinely contribute vital complementary abilities. 

This collaborative model challenges the dominant framing of assistive technology as individualized support, aligning instead with disability studies’ calls to foreground interdependence in design~\cite{bennett_interdependence_2018, branham_collaborative_2015} to support ability-diverse collaboration~\cite{xiao_systematic_2024}, and with work that emphasizes social networks in low-resource contexts~\cite{barbareschi_social_2020}. In these workflows, support is not peripheral—it is structurally embedded. For example, BPS creators frequently rely on sighted collaborators for visual framing and lighting decisions, tasks inaccessible through screen reader interfaces. Similarly, DHH creators may co-produce content with interpreters or hearing collaborators who voice or transcribe signed material. These labour-intensive practices expose limitations in current tooling, which often fails to support such distributed, multimodal work.

Our findings resonate with Xiao et al.'s concept of \textit{ability sharing}~\cite{xiao_systematic_2024}, where collaborators contribute distinct but complementary abilities toward a shared outcome. Creators are not passive recipients of help; they actively shape production decisions, evaluate described visual media, and manage cross-modal communication. Supporting these practices requires designing tools that not only enhance creators' capabilities but also strengthen communication across modalities. For instance, systems could help creators better envision and convey ideas to collaborators and provide accessible real-time feedback on creative outputs. However, unlike high-resource contexts where professional support services may be available, our participants navigated what we term "double digital divides", referring to the intersection of disability-related accessibility barriers with Global South infrastructural constraints.

However, interdependence is often fragile. Many creators rely heavily on personal networks, navigating the constraints of affordability and societal stigma. Without intentional infrastructural support, interdependence risks reinforcing exclusion rather than fostering empowerment.

To address these challenges and better support the collaborative workflows we observed, video creation platforms need fundamental redesigns that acknowledge interdependence as a core feature rather than an accommodation. Based on our findings, we propose three key areas for design intervention:

\begin{itemize}
    \item \textbf{Multimodal Collaborative Interfaces:} Integrate accessible interaction modes (e.g., speech input/output, haptic feedback, visual annotation, and sign language interpretation) to bridge sensory gaps and support mixed-ability collaborations. For example, real-time visual feedback systems could provide audio descriptions of lighting conditions to BPS creators, while DHH creators could benefit from visual cues synchronized with audio elements.
    \item \textbf{Asynchronous and Distributed Collaboration Tools:} Develop cloud-based systems enabling asynchronous task assignment, remote editing, and flexible feedback mechanisms, reducing reliance on immediate local support. These tools should include offline capabilities to address connectivity constraints common in LMIC contexts.
    \item \textbf{Community-driven Support Platforms:} Embed peer-to-peer assistance networks or micro-task crowdsourcing features within content creation platforms, expanding creators' access to dependable support~\cite{teng_help_2024}. Platforms could implement community-based caption verification systems where local users earn micro-payments for providing accessibility support to creators.
\end{itemize}

\subsection{Supporting Control of Agency}
The creative processes of sensory-impaired creators reflect a dynamic interplay between human collaboration and AI assistance. Navigating agency, trust, and risk becomes crucial in both human and AI-mediated interactions. 

Creators often shift between autonomous creation and collaborative engagement, depending on the task. In our study, AI tools were used for (1) description (e.g., Alt-text generation), (2) automation (e.g., sound mastering), and (3) generation (e.g., script drafting). However, creators consistently maintained editorial control, aligning with research positioning AI as a ``catalyst'' rather than a replacement in co-creative systems~\cite{moruzzi_user-centered_2024}.

Multimodal collaboration further complicates agency. Sensory-impaired creators often rely on hybrid communication channels, such as voice input for BPS creators and visual/tactile feedback for DHH creators~\cite{rezwana_designing_2023}. For instance, \textit{AltCanvas}~\cite{lee_altcanvas_2024} uses a dynamic tile-based interface and sonification to support blind users in illustration tasks.

Trust in AI systems is contingent on transparency and explainability. Although participants generally trusted AI for usability and affordability, concerns were raised regarding reliability and cultural misrepresentation. For example, opaque AI-generated captions risked distorting cultural nuances, highlighting the need for explainable interfaces that enable creators to audit and refine outputs~\cite{daly_sensemaking_2025, lin_towards_2024}.

Given these nuanced relationships with both human collaborators and AI tools, design efforts must prioritize creator agency while addressing the specific trust and control concerns our participants articulated:
\begin{itemize}
    \item \textbf{Explainable and Editable AI Outputs:} Provide clear auditing and editing pathways for AI-generated content to ensure alignment with creators’ intentions and cultural contexts~\cite{daly_sensemaking_2025}.
    \item \textbf{Customizable AI Automation:} Allow users to personalize automation features according to sensory preferences and cultural needs, including caption styling and sound mixing.
    \item \textbf{Privacy-sensitive AI Design:} Support offline or local AI processing to protect user privacy, reduce internet dependency, and ensure performance in low-connectivity environments.
\end{itemize}

\subsection{Supporting Authentic Expression within Hegemonic Media Landscapes}
A prominent theme across our interviews was the tension between authentic self-expression and conformity to dominant aesthetic standards. Deaf creators constantly negotiate between using Kenyan Sign Language (KSL) and English, which may reach wider audiences but compromises linguistic authenticity. Similarly, BPS creators expressed concern about meeting conventional production standards that privilege visual aesthetics like precise framing or balanced lighting.

These tensions are not merely technical; they reflect broader dynamics of aesthetic hegemony~\cite{piper_critical_1985}, where dominant standards of production implicitly marginalize creators who operate outside those norms. Rather than compelling creators to conform to inaccessible conventions, design efforts should embrace diverse forms of expression and support flexible storytelling modes. Authenticity, in this context, must be understood as a fluid negotiation shaped by creators’ access, identities, and intended audiences.

Our findings align with Mingus's concept of "access intimacy"~\cite{mia_mingus_access_2017}, wherein accessibility is not merely technical but deeply intertwined with identity, community, and representation. For our participants, authentic expression often involved strategic compromises between representing their disabled identities and meeting audience expectations shaped by ableist norms. As Seo et al. observe in their study of blind YouTubers, creators often feared revealing themselves to the public~\cite{seo_understanding_2021}.

In response, rather than compelling creators to adapt to inaccessible conventions, design efforts should proactively embrace diverse and authentic storytelling through tangible design features. Following Hamraie et al.'s concept of "crip technoscience"~\cite{hamraie_crip_2019}, platforms and tools should explicitly support creators' reimagining and personalization of technologies to reflect their sensory experiences and identities. Specifically, video creation tools could incorporate:
\begin{itemize}
    \item \textbf{Multilingual and Multimodal Captioning:} Enable creators to flexibly toggle or blend captions across local languages, sign languages, and global lingua francas, balancing linguistic authenticity with broader accessibility.
    
    \item \textbf{Adaptive Templates for Non-Visual Editing:} Offer customizable, audio-first editing workflows, including auditory previews, voice-controlled navigation, and tactile feedback, empowering visually impaired creators to independently craft content that reflects their unique aesthetics.
    
    \item \textbf{Expressive Sign Language Interfaces:} Develop specialized editing modes that synchronize sign language narration with visuals and captions, preserving the linguistic authenticity and expressive richness of Deaf creators' content.
    
    \item \textbf{Audience Awareness Features:} Integrate built-in educational elements, such as introductory segments, explanatory tooltips, or interactive overlays, that provide viewers with context about sensory-impaired creators’ production choices and foster broader appreciation of diverse storytelling methods.
\end{itemize}

\subsection{Community Voice, Infrastructural Access, and Knowledge Networks in LMIC Contexts}
Our findings reveal a critical distinction between community-mediated content creation in our study and institutionally supported participatory media documented in ICT4D literature. While Saha et al.'s Benefits of Community Voice (BoCV) framework \cite{saha_community_2023,saha_hearing_2024} demonstrates how marginalized creators, specifically women farmers in Bangladesh, leverage participatory videos to amplify voices and influence international program development, our participants operated largely without such institutional mediation.
Where Saha et al. document NGO-facilitated content creation that serves advocacy and policy influence purposes, our Kenyan creators with sensory impairments relied primarily on informal peer networks and family support systems. This contrast highlights a significant gap in institutional support for disabled creators in LMIC contexts. Our participants' content served multiple purposes, including livelihood generation, disability awareness, artistic expression, and community building, yet lacked the structured advocacy pathways available to other marginalized groups. 
This reliance on informal networks reflects broader infrastructural gaps that our study revealed. Only four of our twenty participants had formal training in video production: three DHH creators who attended a government-supported training initiative, and one BPS creator who joined a short photography course sponsored by Canon. For most, entry into content creation was driven by personal initiative, informal peer learning, and support from collaborators. While many creators demonstrated creative capacity and a strong drive to share their stories, they lacked access to reliable equipment, affordable internet, and formal training opportunities.
Their participation was made possible not through systemic inclusion but through ad hoc solutions and community knowledge exchange. This reliance on informal networks, while demonstrating remarkable community resilience, also exposes creators to greater vulnerability. Unlike the sustained institutional support documented in ICT4D participatory media projects, our participants faced precarious dependencies on unpaid assistance, highlighting the need for more robust support infrastructures that bridge individual creativity with community empowerment.
These findings align with global research indicating that marginalized creators often navigate layered exclusions, including technical, economic, and social barriers. Tsatsou's qualitative study [54] highlights the intra-disability diversity in digital inclusion, emphasizing the role of individuality and selectiveness in digital engagement among people with disabilities. Hunt et al.'s review [26] identifies persistent gaps in access to community support services, the need for stronger infrastructural systems, and the critical role of community-driven strategies like peer support and informal mentoring. It emphasizes that without investment in formal community-based support structures, such as training, equipment access, and accessible digital platforms, participation risks remaining unsustainable and precarious. While inclusive tool design can address some of these challenges, systemic support structures (e.g., training programs, mentorship, institutional funding) are essential for enabling long-term, sustainable participation in digital content ecosystems. However, inclusive tools can significantly mitigate these gaps through targeted design:
\begin{itemize}
    \item \textbf{Compatibility with Affordable Hardware:} Ensure reliable performance on entry-level smartphones and basic laptops common in LMICs. This includes optimizing applications for devices with limited RAM and processing power.
    \item \textbf{Integrated Digital Literacy Support:} Embed interactive, language-appropriate digital skill-building modules directly into creation platforms. These could include step-by-step tutorials available in Swahili and Kenyan Sign Language, with accessibility features for both BPS and DHH learners.
    \item \textbf{Strengthening Community Networks:} Provide integrated forums and mentoring features to support peer learning and collaborative problem-solving. Drawing from successful community-driven models in ICT4D, platforms could implement structured peer mentorship programs that connect experienced creators with newcomers, facilitating knowledge transfer within the disability community.

\end{itemize}

%% file: Chapters/7_Conclusion.tex
\section{CONCLUSION}

We conducted a qualitative study with 20 video content creators with visual and hearing impairments in Kenya to investigate their tools, challenges, practices, and collaborative strategies. 
We examined how accessibility barriers intersect with infrastructural limitations, shaping content creation as a collaborative, adaptive, and interdependent practice.

Our participants navigated challenges that extended beyond universal accessibility barriers to include constraints specific to LMIC contexts. While BPS creators struggled with camera framing and visual interface navigation, and DHH creators faced difficulties with audio editing and captioning accuracy, these technical challenges were amplified by financial barriers, device limitations, absent training opportunities, language barriers, and platform monetization challenges that created "double digital divides" for disabled creators in resource-constrained environments.

We found that video content creation among sensory-impaired creators was inherently a staged and collaborative process, distributed across pre-production, filming, editing, and post-production. This process involved trusted human collaborators as well as emerging AI tools. Creators negotiated agency and trust throughout their workflows, actively managing the tension between seeking assistance to bridge sensory gaps and maintaining creative control over their work.

Drawing from these insights, we argue that accessible video creation for sensory-impaired creators necessitates supporting interdependent collaboration, fostering flexible human-AI workflows, and embracing diverse storytelling practices. We highlight the importance of design approaches that prioritize agency, trust, and authentic expression, while addressing the infrastructural and social barriers that constrain disabled creators.

As one of the first studies to explore end-to-end video creation workflows of sensory-impaired creators in a low-resource urban context, our work expands the sociotechnical scope of accessibility research within HCI. We call for the development of inclusive systems that not only accommodate but also amplify the creativity and agency of disabled creators globally.